\documentclass[journal]{IEEEtran}

\IEEEoverridecommandlockouts 

\usepackage{amsfonts}
\usepackage{graphicx}
\usepackage{cite}

\input{steve.lib}

\title{A Distributed Observer for a Time-Invariant Linear System}

\author{L. Wang$^{1}$, and A. S. Morse$^{1}$
\thanks{*This work was  supported by  National Science Foundation grant n. 1607101.00 and  US Air Force grant n. FA9550-16-1-0290.}
\thanks{$^{1}$ L. Wang and A. S. Morse are with the Department
of Electrical Engineering, Yale University, New Haven, CT, USA.
        {\tt\small \{lili.wang,  as.morse\}@yale.edu }}%
        }

\begin{document}

 \maketitle
 \thispagestyle{empty}

\begin{abstract} A   time-invariant, linear, distributed  observer is described for  estimating the state
of an $m>0$ channel, $n$-dimensional continuous-time linear system
of the form $\dot{x} = Ax,\;y_i = C_ix,\;i\in\{1,2,\ldots, m\}$. The
state $x$ is   simultaneously estimated by $m$ agents assuming each
agent $i$ senses  $y_i$ and receives the state $z_j$ of each of its
neighbors' estimators. Neighbor relations are characterized by a
constant directed  graph $\mathbb{N}$ whose
 vertices correspond to agents and whose arcs depict neighbor relations. For the case when the neighbor graph is strongly connected, the overall distributed observer consists of
 $m$ linear estimators, one for each agent; $m-1$ of the estimators are of dimension  $n$ and one estimator is of dimension $n+m-1$.
  Using  results from classical decentralized control theory, it is shown that subject to  the assumptions that
   (i) none of the $C_i$ are zero,
(ii) the   neighbor graph $\mathbb{N}$ is strongly connected, (iii)
the system whose state is to be estimated is jointly observable,
 and nothing more,
    it is possible to freely assign the spectrum of the overall distributed observer.  For the  more general
    case when $\mathbb{N}$ has $q>1$ strongly connected components,
    it is  explained how to construct a family of  $q$ distributed
    observers, one for each component, which can estimate $x$ at a preassigned convergence rate.
\end{abstract}

\section{Introduction}

State estimators such as Kalman filters and observers have had a
huge impact on the entire field of estimation and control. This
paper deals with observers for time-invariant linear systems.
 An observer for a process modeled by a continuous-time,  time-invariant  linear system with state $x$,    measured output
  $ y =Cx$ and
 state-dynamics $\dot{x} = Ax$, is a  time-invariant  linear system  with input $y$ which is capable
 of generating an asymptotically correct estimate of $x$ exponentially fast at a pre-assigned but arbitrarily large
  convergence rate.  As is well known,  the only requirement on the system $ y =Cx$, $\dot{x} = Ax$ for such
  an estimator to
exist is that the matrix pair $(C,A)$ be observable. In this paper
we will be  interested in the
 natural generalization of this
 concept appropriate to a   network of $m$ agent.  We now make precise  what we mean by this.

\subsection{The Problem}
We are interested in a  fixed network of $m>0$ autonomous agents
labeled $1,2,\ldots, m$  which are able to receive information from
their neighbors where by the  {\em neighbor} of agent $i$ is meant
any other agent in agent $i$'s reception range. We write
 $\scr{N}_i$ for the set of labels of agent $i$'s neighbors and we take agent $i$ to be a neighbor of itself.
  Neighbor relations between distinct  pairs of agents are
 characterized  by a directed graph $\mathbb{N}$ with $m$ vertices and a set of arcs defined so that there is an arc from
  vertex $j$ to vertex $i$ if whenever agent $j$ is a  distinct neighbor of agent $i$. Each agent $i$
  can sense a signal $y_i\in\R^{s_i},\;i\in\mathbf{m}=\{1,2,\ldots, m\}$, where \begin{eqnarray}y_i &=&C_ix,\;\;\;i\in
  \mathbf{m}\label{sys1}\\\dot{x} &= &Ax\label{sys2}
\end{eqnarray}
and $x\in\R^n$.

Agent $i$ estimates $x$ using an $n_i$ dimensional  linear system
with state vector $z_i$   and we assume the information agent $i$
can receive from neighbor $j\in\scr{N}_i$ is $z_j(t)$ and $y_j(t)$. 
The problem of interest is to construct a suitably defined family of
linear systems
\begin{eqnarray}
\dot{z}_i & = & \sum_{j\in\scr{N}_i}(H_{ij}z_j  + K_{ij}y_j),\;\;\;i\in\mathbf{m}\label{ob1}\\
x_i & = & \sum_{j\in\scr{N}_i}(M_{ij}z_j  +
N_{ij}y_j),\;\;i\in\mathbf{m}\label{ob2}\end{eqnarray}
 in such a way so that no matter what the initializations of \rep{sys1} and \rep{ob1},  each signal $x_i(t)$ is
  an asymptotically correct estimate of $x(t)$ in the sense that each
   {\em estimation error} $e_i = x_i(t)-x(t)$ converges to zero as $t\rightarrow \infty$ at a preassigned, but arbitrarily fast convergence rate.
    We call such a family a {\em distributed (state) observer}.

    We assume throughout that $C_i \neq 0,\;i\in\mathbf{m}$, and that the system defined by \rep{sys1}, \rep{sys2} is {\em jointly observable};
     i.e.,  with
    $C = \matt{C_1' &C_2' &\cdots &C_m'}'$, the matrix pair $(C,A)$ is observable. Generalizing the results which follow to the case when
    $(C,A)$ is only detectable is quite straightforward and can be accomplished using well-known ideas.
    {However the commonly made assumption
      that each pair
    $(C_i,A),\;i\in\mathbf{m}$, is observable, or even just detectable, is very restrictive, grossly
    simplifies the problem and is as unnecessary. It is precisely the exclusion of this assumption  which distinguished
    the problem posed  here from
    almost all of the distributed estimator problems addressed in the literature.  The one exception we are aware of is
    the recent paper \cite{martins} which has provided the main motivation for this work.

\subsection{Background}
There is a huge literature  which seeks to deal with distributed
Kalman filters or distributed observers; see, for example
\cite{martins,shamma,carli,xxx,saber2,bullo.observe} and the many
references cited therein. Many result  are only partial and most
problem formulations are different in detail than the problem posed
here.
 The problem we have posed was prompted  specifically by the
  work in \cite{martins} which seeks to devise a
 time-invariant distributed observer for the  discrete-time analog of \rep{sys1}, \rep{sys2}. Two particularly important contributions are made
 in \cite{martins}.  First it is recognized
 that the problem of crafting a `stable'
 distributed observer  is more or less equivalent to  devising a stabilizing decentralized control as in \cite{wangdavison,corfmat}.
Second, it is demonstrated that under suitable conditions, it is
only necessary for the dimension of one of the agent subsystems in
\cite{martins} to be larger than $n$, and that the enlarged
dimension need not exceed $n+m-1$.

The work reported in this paper clarifies and expand on the results
of \cite{martins} in several ways. First we outline a construction
for systems with strongly connected
 neighbor graphs  which enables one to freely adjust the observer's spectrum. Second,
 the results obtained  here apply whether $A$ is singular or not; the implication of this generalization is that  the construction
 proposed  can be used to craft observers for continuous time processes whereas the construction proposed in
 \cite{martins} cannot unless $A$ is nonsingular.

 \section{Observer Design Equations}

We now develop the interrelationships between the matrices appearing
in \rep{ob1} and \rep{ob2} which must hold for each $x_i$ to be
 an asymptotically
 correct estimate of $x$.
Note first that because \rep{ob2} must hold  even when all
 estimates are correct, for each $i\in\mathbf{m} $ it is necessary that the  equation
$x =  \sum_{j\in\scr{N}_i}(M_{ij}z_j  +
N_{ij}C_jx),\;\;i\in\mathbf{m}$
 have a solution $z_{i}^x,\;i\in\mathbf{m},$ for each possible $x\in\R^n$.
Thus if we define $V_i = \matt{z_{i}^{u_1} &z_{i}^{u_2} & \cdots &
z_{i}^{u_n}}_{n_i\times n},\;i\in\mathbf{m}$,
 where $u_k$ is the $k$th unit
vector in $\R^n$, then \eq{\boxed{I =\sum_{j\in\scr{N}_i}
(M_{ij}V_{j} +N_{ij}C_j),\;\;\;i\in\mathbf{m}}\label{design1}} This
and \rep{ob2}  imply that the $m$ estimation errors  satisfy
\eq{x_i-x = \sum_{j\in\scr{N}_i}M_{ij}\epsilon_j,\;\;i\in\mathbf{m}
\label{esterror}} where \eq{\epsilon_i = z_i -
V_ix,\;i\in\mathbf{m}\label{interror}} Moreover, as a direct
consequence of \rep{sys1}, \rep{sys2}, and \rep{ob1},
\[\dot{\epsilon}_i = \sum_{j\in\scr{N}_i}H_{ij}\epsilon_j +\left
(\sum_{j\in\scr{N}_i}(H_{ij}V_j  +K_{ij}C_j)-V_iA\right
)x,\;\;i\in\mathbf{m}\] Thus if we stipulate that \eq{\boxed{V_{i}A
=\sum_{j\in\scr{N}_i}(H_{ij}V_{j}
+K_{ij}C_j),\;i\in\mathbf{m}}\label{design2}}
 then
\eq{\dot{\epsilon}_i =
\sum_{j\in\scr{N}_i}H_{ij}\epsilon_j,\;i\in\mathbf{m}\label{int}} We
shall refer to \rep{design1} and\rep{design2} as the {\em observer
design equations}. These equations are quite general. They apply to
all
 time-invariant continuous and discrete time state observers whether they are distributed or not.

  It is clear from \rep{esterror} that if the $V_i, H_{ij}, M_{ij}, N_{ij}$ and $K_{ij}$
can be chosen so that the observer design equations  \rep{design1},
\rep{design2} hold and  the system defined by \rep{int} is
exponentially stable, then each $x_i$ will be an asymptotically
correct estimate of $x$. The {\em distributed observer design
problem} is to develop constructive conditions which ensure that the
$V_i, H_{ij}, M_{ij}, N_{ij}$ and $K_{ij}$ can be so chosen.

\section{Centralized Observers}

The purpose of this section is to review the  well-known concept of
a (centralized) observer with the aim summarizing certain less well
know ideas which will play a role in the construction of a
distributed observer. In the centralized case $m=1$ and a state
observer is a $n_1$-dimensional linear system with input $y=Cx$,
state $z\in\R^{n_1}$ and output $x_1$ of the form $\dot{z}=Hz+Ky$,
$x_1=Mz+Ny$. In this case the observer design equations are
$I=MV+NC$ and $VA=HV+KC$ and the observer design problem  is to
determine  matrices $H,K,M,N$ and $V$
  so that the observer design equations hold and $H$ is a stability matrix.
 Observers fall into three broad categories depending
   on the dimension $n_1$: full state observers, minimal state observers,  and extended state observers. Each type is briefly reviewed below.

\noindent{\bf Full-State Observers:} Just about the easiest solution
to the observer design problem that one can think of, is the one for
which $M\dfb I$, $N=0$, $V=I_{n\times n}$ and $H\dfb A-KC $. Any
observer of this type is called a {\em full-state observer} because
in this case $z_1$ is an asymptotic estimate of $ x$.
  Of course  it is necessary  that $K$ be chosen so that $A-KC$ is a
stable matrix.
 One way to accomplish this is
to exploit duality and use spectrum assignment, as is well known. No
matter how one goes about defining $K$, the
 definitions of $H,M,N,$ and $V$ given above show that a full-state observer is modeled by equations of the form
$\dot{z}_1=(A-KC)z_1+Ky$, $x_1=z_1$.

\noindent{\bf Reduced State Observers:}
  By a {\em minimal state observer} is meant an observer of least dimension which
  can generate an asymptotic estimate of $x$.
  Minimal dimensional observers are obtained by exploiting the fact that $y=Cx$ is a ``partial'' measurement of $x$.
   Note that the observer design equation $I=MV+NC$
   implies that the number of linearly independent rows of  $V_{n_1\times n}$ must be
 at least equal to the dimension of  $\ker C$.  Thus the dimension of any  observer must be at least equal to dimension
 $\ker C$.  Techniques for constructing minimal state observers are well known \cite{luenberger,wonham.observe}.


\noindent{\bf Extended State Observers:}  Much less well known are
what might be called `extended state observers.'  A  observer of
this type would be of dimension $n_1 = n +\bar{n}$
 where $\bar{n}$ is
 a nonnegative integer chosen by the designer.  With $\bar{n}$ fixed, an extended observer can be obtained  by first
 picking $M = \matt{I & 0}_{n\times (n+\bar{n})}, V' = \matt{I & 0}_{n\times (n+\bar{n})} $ and $N = 0$ thereby
 ensuring that observer design equation $I=MV+NC$
 is satisfied. With $V$ so chosen, $z_1$ must be of the corresponding
  form $z_1=\matt{x_1' &\bar{z}_1'}'$.  Accordingly, the partitioned matrices
$$H = \matt{A+\bar{D}C & \bar{C}\cr \bar{B}C & \bar{A}}_{(n+\bar{n})\times (n+\bar{n})}\hspace{.4in}K = -\matt{\bar{D}\cr \bar{B}} $$
satisfy the observer design equation $VA=HV+KC$ for any values of
the matrices $\bar{A}, \bar{B}, \bar{C}, \bar{D}$ and
\begin{eqnarray*}
\dot{{x}}_1 & =  & (A+\bar{D}C)x_1 +\bar{C}\bar{z}_1 -\bar{D}y\\
 \dot{\bar{z}}_1 & = & \bar{B}Cx_1 +\bar{A}\bar{z}_1 - \bar{B}y\end{eqnarray*}
 Moreover the estimation error $e=x_1-x$ satisfies
 \begin{eqnarray*}
 \dot{e} &=& (A+\bar{D}C)e +\bar{C}\bar{z}_1 \\
 \dot{\bar{z}}_1 &=& \bar{B}Ce +\bar{A}\bar{z}_1
 \end{eqnarray*}
These equations suggests the following feedback diagram.
\begin{figure}[h]
\centerline{\includegraphics[height = 1.2in]{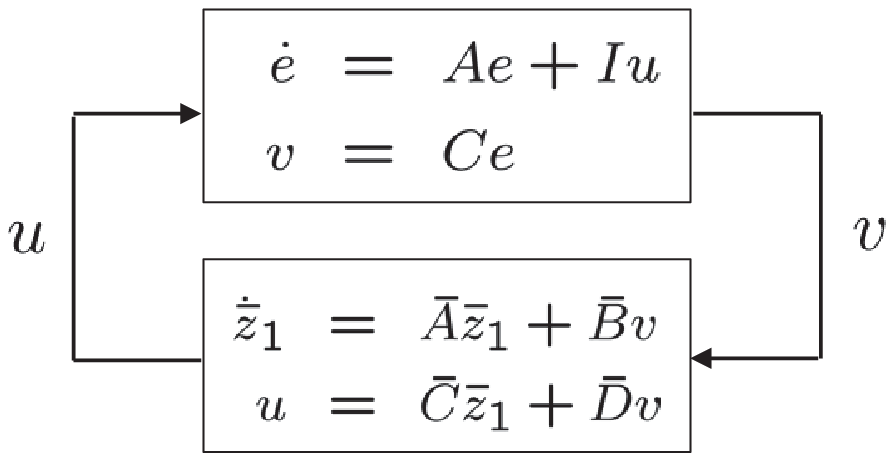}}
\label{feeder}\end{figure}

\noindent Thus the design of an extended state observer amounts  to
picking the coefficient
 matrices $\{\bar{A},\bar{B},\bar{C},\bar{D}\}$ of the lower subsystem
 in the block diagram to at least stabilize the loop. Of course if $\bar{n} = 0$, this subsystem  is just the constant matrix $\bar{D}$
 and one has again a classical full-state  observer of dimension $n$.  Exactly what might be gained by picking $\bar{n}$
  greater than zero  is not clear in the case of a centralized  observer.  However, for the decentralized observer we describe next,
  the flexibility of a dynamic lower loop will become self-evident.

\section{Distributed Observers}

The primary goal of distributed observer design is to choose the
matrices
 $V_i, H_{ij}, M_{ij}, N_{ij}$ and $K_{ij}$
 so that the observer design equations \rep{design1} and \rep{design2}
hold and  the system defined by \rep{int} is exponentially stable.
Another goal might be to choose these matrices to reduce the
information which needs to be  transmitted between neighboring
agents.  Still another goal might be to choose these matrices so
that the dimensions of the individual estimators are as small as
possible.  In this paper we will consider the case when the only
information transmitted between neighboring agents are estimator
states $z_i$ and we will make no attempt to construct estimators of
least dimension.  This means that  we will set all  $K_{ij} = 0$
except for $K_{ii}$ in \rep{ob1} and all
 $N_{ij} = 0$ in  \rep{ob2}.
The easiest way to satisfy the observer design equations is to  set
 $V_i = I_{n\times n}$ for $\;i\in\mathbf{m}$ and to pick the $M_{ij}$ so that
$I = \sum_{j\in\scr{N}_i}M_{ij} ,\;i\in\mathbf{m}$.  With the $V_i$
so chosen, observer design equation \rep{design2} simplifies to
\eq{A-K_iC_i =\sum_{j\in\scr{N}_i}H_{ij}
,\;i\in\mathbf{m}\label{form1}} where we have adopted the notation
$K_i = K_{ii}$ In view of \rep{esterror} and \rep{int}, the observer
design problem for this type of an observer is to try to choose the
 $K_i$ and $H_{ij}$ so that \rep{form1} holds
 and in addition so that
$H=\matt{H_{ij}}$ is a stability matrix where $H_{ij} = 0$ if
$j\not\in\scr{N}_i$.
 It is possible to express $H$ in a more explicit form which takes into account the constraints on the $H_{ij}$ imposed by \rep{form1}.
 For this
let $\tilde{A}$ denote the block diagonal matrix $\tilde{A} =
I_{m\times m}\otimes A $ where $\otimes $ is the  Kronecker product.
Set $B_i = b_i\otimes I_{n\times n}\;i\in\mathbf{m}$ where $b_i$ is
the $i$th unit vector in $\R^m$;
 in addition,  let
  $C_{ii} = C_iB_i',\;i\in\mathbf{m}$, and
 $C_{ij} = c_{ij}\otimes I_{n\times n},\;\;j\in\scr{N}_i,j\neq i,\;i\in\mathbf{m}$
where $c_{ij}$ is the row in the transpose of the incidence matrix
of $\mathbb{N}$ corresponding to the arc  from $j$ to $i$. It is
then
 possible to express
$H$  in the compact form \eq{H = \tilde{A}+
\sum_{i\in\mathbf{m}}\sum_{j\in\scr{N}_i}B_iF_{ij}C_{ij}\label{H}}
where $F_{ii} = -K_i,i\in\mathbf{m}$ and $F_{ij} =
H_{ij},\;j\in\scr{N}_i,j\neq i,\;i\in\mathbf{m}$. Note that there
are no
 constraints on the  $F_{ij}$. In this form it is clear that $H$ is what results when output feedback laws $u_{ij} = F_{ij}y_{ij}$
 are applied to the system
 \begin{eqnarray}\dot{\epsilon} & = &\tilde{A}\epsilon + \sum_{i\in\mathbf{m}}\sum_{j\in\scr{N}_i}B_iu_{ij}\label{decent}\\
y_{ij}& =&
C_{ij}\epsilon,\;\;\;\;ij\in\scr{I}\label{decent2}\end{eqnarray}
where $\scr{I}\subset \mathbf{m}\times \mathbf{m}$  is the set of
double indices $\scr{I} = \{ij:i\in\mathbf{m},\;j\in \scr{N}_i\}$.
The problem of constructing a distributed observer of this type thus
reduces to trying to  choose the $F_{ij}$ to at least stabilize $H$
 if such matrices exist. Of course, one also wants control over rate of convergence, so stabilization  of $H$ alone is not all that is
  of interest. Whether the goal is just stabilization of $H$ or control over convergence rate, choosing the $F_{ij}$ to accomplish this will
   typically not be possible
  except under special conditions. In fact the problem trying to stabilize $H$ by appropriately choosing the $F_{ij}$
  is mathematically the same as the   classical decentralized stabilization problem
 for which there is a substantial literature  \cite{wangdavison,corfmat}.

\subsection{Strongly Connected Neighbor Graph $\mathbb{N}$}

  One approach is to decentralized stabilization problem  is
 to try to choose the $F_{ij}$ so that for  given $p\in\mathbf{m}$ and $q\in\scr{N}_p$,
 the matrix pairs $(H,B_p)$ and $(C_{pq},H)$ are controllable and observable respectively.
Having accomplished this, stabilization can then be achieved by
applying standard centralized feedback
 techniques such as those in \cite{braschpearson}
to the resulting controllable observable system. This is the
approach taken in this paper. The following proposition provides the
key technical result  which we need.

\begin{proposition} Suppose that the neighbor graph $\mathbb{N}$ is strongly connected.
 There exist gain matrices $F_{ij},\;ij\in\scr{I}$ such that  the matrix pairs $(H,B_p)$ and $(C_{pq},H)$ are
 controllable and observable respectively for all $p\in\mathbf{m}$ and all $q\in\scr{N}_p$. Moreover, for any such pair,
 $m$ is the controllability index of $(H,B_p)$. \label{prop1}\end{proposition}
The proof of this proposition
will be given is Section \ref{sec:analysis}.

In the light of Proposition \ref{prop1}, the way to construct a
distributed observer is clear. As a first step, choose matrices
$M_{ij}$, $i\in\mathbf{m},\;j\in\scr{N}_i$ so that $I =
\sum_{j\in\scr{N}_i}M_{ij} ,\;i\in\mathbf{m}$. Next choose the
$F_{pq}$ so that the conclusions of the Proposition \ref{prop1}
hold.
 Having so chosen the $F_{ij}$ or equivalently the $H_{ij}$ and the $K_i$, fix values of $p\in\mathbf{m}$ and $q\in\scr{N}_p$.
Next set $\bar{n} = m-1$ and
 use a standard construction technique such as that given in \cite{braschpearson} to pick   matrices
$ \bar{A}_{\bar{n}\times \bar{n}}, \bar{B}_{\bar{n}\times \omega},
\bar{C}_{n\times \bar{n}}$ and $\bar{D}_{n\times \omega}$ to assign
a desirable  spectrum to the matrix
$$\bar{H} = \matt{H +B_p\bar{D}C_{pq} &B_p\bar{C}\cr  \bar{B}C_{pq}  & \bar{A}}_{(nm+\bar{n})\times (nm+\bar{n})}$$
where  $\omega = s_p$ if $q=p$ or
  $\omega = n$ if $p\neq q$.
This can be done because $(C_{pq}, H)$ is an observable pair  and
because $(H,B_p)$ is a controllable pair with controllability index
$m$. The corresponding distributed observer equations are
\begin{eqnarray*}
\dot{z}_i & = & \sum_{k\in\scr{N}_i}H_{ik}z_k  + K_iy_i,\;\;\;i\in \mathbf{m}, i\neq p \\
\dot{z}_p & = & \sum_{k\in\scr{N}_p}H_{pk}z_k  + K_py_p +\bar{C}\bar{z} + \bar{D}v \\
\dot{\bar{z}} &=& \bar{A}\bar{z} + \bar{B}v\\
x_i & = & \sum_{k\in\scr{N}_i}M_{ik}z_k,  \;\;i\in\mathbf{m}\\
\end{eqnarray*}
where $v= y_p $   if $q=p$ or
 $v = x_p-x_q$  if $p\neq q$.

 It is possible to verify that the observer design equations hold. For simplicity, assume that $p=m$ and
  redefine $V_m$ to be $\matt{I & 0}'_{(n+\bar{n})\times n}$.
 For $k\in\mathbf{m}$, redefine $M_{km}$
to be $\matt{M_{km} & 0}_{\omega \times (n+\bar{n})}$ thereby
ensuring that  observer design equation \rep{design1} holds. To
ensure that observer design equation \rep{design2} holds, first
replace $H_{km}$ with $\matt{H_{km} & 0}_{n\times (n+\bar{n})}$ for
$k\in\{1,2,\ldots,m-1\}$. If $q=m$ replace  $H_{mi
},\;i\in\mathbf{m},\; i\neq m$, and $H_{mm}$   with the matrices
$$\matt{H_{mi} \cr  0},i\in\{1,2,\ldots, m-1\}\hspace{.2in}$$  and
$$  \matt{ H_{mm} +\bar{D}C_m &\bar{C}\cr \bar{B}C_m &\bar{A}}$$
respectively; in addition, replace $K_m$ with the matrix $
\matt{(K_m-\bar{D})'& -\bar{B}'}'_{(n+\bar{n})\times s_m}$. If on
the other hand, $m\neq q$, replace  $H_{mi },\; i\in
\mathbf{m},\;i\neq p,m$, $H_{mp}$ and $H_{mm}$  with the matrices
$$\matt{H_{mi}  \cr  0}, i\in \{1,2,\ldots,m-1\}, i\neq q,$$
$$ \matt{H_{mq} -\bar{D} \cr  -\bar{B}}, \hspace{.2in}  \rm{and } \matt{ H_{mm} +\bar{D} &\bar{C}\cr
\bar{B} &\bar{A}}$$ respectively;\ \ in addition, \ \ replace $K_m$
with matrix $\matt{K_m' & 0}'_{(n+\bar{n})\times s_m}$. In either
case, observer design equation  \rep{design2} holds. Thus the
 error $\bar{\epsilon}  = \matt{\epsilon_1' &\epsilon_2' & \ldots &\epsilon_m'  &\bar{z}'}'$
 satisfies $\dot{\bar{\epsilon}} = \bar{H}\bar{\epsilon}$ where as before,
$x_p-x = \sum_{k\in\scr{N}_p}M_{pk}\epsilon_p,\;\;p\in\mathbf{m} $.

We are led to the main result of this paper.

\begin{theorem} Suppose that \rep{sys1}, \rep{sys2} is a jointly observable system and that $C_i\neq 0,\;i\in\mathbf{m}$.  If
 the neighbor graph $\mathbb{N}$ is strongly connected, then for each symmetric
 set of $mn +m-1$ complex numbers $\Lambda$ there is a distributed  observer \rep{ob1}, \rep{ob2} for which the spectrum of
the  $(mn +m-1) \times (mn +m-1)   $ matrix $H \dfb \matt{H_{ij}}$
is  $\Lambda $.  Moreover, the observer's  $m$  outputs
$x_i(t),\;i\in\mathbf{m}$, all
   asymptotically correctly estimate $x(t)$ in the sense that each
   estimation error $e_i = x_i(t)-x(t)$ converges to zero as $t\rightarrow \infty$ as fast $e^{Ht}$ converges to zero,
 no matter what the initializations of \rep{sys2} and \rep{ob1} are.
    \label{main} \end{theorem}

\subsection{ Non-Strongly Connected  Neighbor Graph $\mathbb{N}$}

We now turn briefly to the problem of developing a distributed
observer for
 the case when $\mathbb{N}$ is not strongly connected. We will assume for simplicity and
  without loss of generality that $\mathbb{N}$ is weakly connected.  For if it is not,  the ideas which follow can be applied  to each
  maximally  weakly connected subgraph of $\mathbb{N}$, since each such subgraph is  isolated  from the rest.
As before, the goal is to devise  $m$ estimators whose estimates
 converge to $x$ exponentially fast at arbitrary, pre-assigned rates.
We suppose that $\mathbb{N}$ has $q$ strongly connected components
  $\mathbb{N}_1,\mathbb{N}_2,\ldots,\mathbb{N}_q$ and for each $i\in\mathbf{q}$ we write $\Sigma_i$ for the $m_i$ channel
  {\em component subsystem}
    $\dot{x} = Ax$ $y_j = C_jx,\;j\in\scr{V}_i$  where $\scr{V}_i$ is the set of labels of
  the vertices of $\mathbb{N}_i$ and $m_i$ is the number of labels in $\scr{V}_i$.
 We say that there is a directed  path \{resp. arc\} from  strongly connected component
  $\mathbb{N}_i$ to strongly connected component $\mathbb{N}_j$
if there is a directed path \{resp. arc\} in $\mathbb{N}$ from at
least one vertex in $\mathbb{N}_i$ to at least one vertex in
$\mathbb{N}_j$.
 Following \cite{martins}, we say that $\mathbb{N}_j$ is a {\em source component} of $\mathbb{N}$ if
  $\mathbb{N}_j$ has no incoming arcs from any other strongly connected component of $\mathbb{N}$. It is clear that $\mathbb{N}$ must contain at
  at least one source component. Moreover, since $\mathbb{N}$ is weakly connected, it is also clear that
   for any strongly connected component
  of $\mathbb{N}_i$  which is not a source, there must be at least one directed path from  at least one source $\mathbb{N}_j$  to
$\mathbb{N}_i$.

Let $\mathbb{N}_j$ be a source component and $\Sigma_j$ be its
associated  component subsystem.
 Note that there cannot be any signal
flow to any channel in $\Sigma_j$ from any  channel of any other
component subsystem.   It follows that
  for there to exist
 estimators  for each channel in $\Sigma_j$  which are  capable of estimating $x$ at a preassigned convergence rate, it is  necessary
 that   $\Sigma_j$  be  a jointly observable subsystem. In view of Theorem \ref{main},
 joint observability of $\Sigma_j$ is  also sufficient for such a distributed observer to exist because $\mathbb{N}_j$ is strongly connected.
  Suppose therefore that for each source component $\mathbb{N}_j$,
the associated component subsystem $\Sigma_j$ is jointly observable
and that a distributed  observer has  been constructed with
preassigned converge rate for each such  $\Sigma_j$. If all strongly
connected components of $\mathbb{N}$ are sources, then these
observers solve the distribute observer design problem.
  Suppose therefore that
there is at least one
 strongly connected component   which is not a source.
 Then there must be at least one strongly connected component $\mathbb{N}_i$ which is not a source for which there
 is a source $\mathbb{N}_j$ with an arc to
  $\mathbb{N}_i$.
 This implies that  there must be a channel $k\in\scr{V}_j$  of  $\Sigma_j$
  whose  estimator  state $z_k$  is available to at least one channel - say channel $l$  of component subsystem $\Sigma_i$.
   But $\epsilon_k = z_k - V_kx$. Moreover, for the full-state observers we are considering, $ V_k'$  is a left inverse of $V_k$ so
   $ V_k'z_k = \bar{C}_lx + V_k'\epsilon_k$  where $\bar{C}_l = I_{n\times n}$. Therefore  $V_k'z_k$ can be regarded as a
   measurement of  $x$ with  exponentially decaying additive  measurement noise  $V_k'\epsilon_k$. Thus if the readout equation $y_l = C_lx$
   in the definition of $\Sigma_i$, is replaced with with the augmented readout equation
$$y_l = \matt{C_l\cr \bar{C}_k}x +\matt{0\cr V_k'\epsilon_k},$$ then the resulting subsystem, denoted by $\bar{\Sigma}_i$
 will be  jointly observable with unmeasurable but exponentially decaying measurement noise.
 Since
   $\mathbb{N}_i$ is strongly connected, a distributed observer with  the same convergent rate  as that of $\epsilon_k$, can therefore
    be constructed
   for  $\Sigma_i$.
    If $\mathbb{N}_i$ is the only strongly connected component of $\mathbb{N}$ which is not a source, then construction is complete.
    If, on the other hand, $\mathbb{N}$
     has other strongly connected components which are not sources, the same ideas as just described, can be applied to each
     corresponding component subsystem in a sequential manner.
      We are led to the following

\begin{corollary} Suppose that $C_i\neq 0,\;i\in\mathbf{m}$ and  that neighbor graph $\mathbb{N}$ has $q$ strongly connected components
$\mathbb{N}_i,\;i\in\mathbf{q}$.
 Let
 $\Sigma_i$ be the component subsystem of \rep{sys1}, \rep{sys2} corresponding to strongly connected component $i$.
  In order for there to exist distributed observers for each of the component subsystems which are a capable of estimating $x$ at an
  arbitrary but preassigned convergence rate, it is necessary and sufficient that each of the component subsystems whose graphs are sources,
  are jointly observable.
\end{corollary}

\section{Decentralized Control Theory}

The aim of this section is to summarize the concepts and results
from \cite{corfmat} and \cite{corf} which we will make use of to
justify Proposition \ref{prop1}. We do this for a  $k$ channel,
$n$-dimensional linear system of the form \eq{\dot{x} = Ax
+\sum_{i\in\scr{I}}B_iu_i\hspace{.5in} y_i =
C_ix,\;\;i\in\scr{I}\label{generic}} where
 $\scr{I} = \{1,2,\ldots, k\}$\footnote{ The symbols used in this section such as
$x, C_i,A,\scr{I}$ are generic and do not have the same meanings as
the same symbols do  when used elsewhere in the paper.}
 and $C_i \neq 0,\;i\in\scr{I}$.
Application of decentralized feedback laws  of the form $u_i =
F_iy_i,\;i\in\scr{I}$ to this system yields the equation $\dot{x} =
H x$ where
 $H = A + \sum_{i\in\scr{I}}B_iF_iC_i$.  For given $p\in\scr{I}$, explicit necessary and sufficient conditions
 under which there exist $F_{i}$
  which make
 $(C_{p},H,B_p)$ controllable and  observable  are given in\cite{corf} and \cite{corfmat}.
   There are two conditions.
First, \rep{generic} must be jointly controllable and jointly
observable. Second, each ``complementary subsystem'' of
  \rep{generic} must be ``complete.'' \{cf. Theorem 3, \cite{corfmat}\}
 There are as many complementary subsystems
 of \rep{generic} as there are strictly proper subsets of  $\scr{I}$.
 By the {\em complementary  subsystem} of \rep{generic}
 corresponding to a nonempty  proper subset $\scr{C}\subset \scr{I}$, is meant a subsystem with input matrix
 $\mathbf{B}(\scr{C}) = {\rm block\;row}\{B_i:i\in\scr{C}\}$, state matrix $A$ and readout matrix
 $\mathbf{C}(\bar{\scr{C}}) = {\rm block\;column}\{C_{i}:i\in\bar{\scr{C}}\}$
where $\bar{\scr{C}}$ is the complement of $\scr{C}$ in $\scr{I}$
\cite{corfmat}. The complementary subsystem determined by $\scr{C}$
is
 uniquely determined up to the orderings of
 the block rows and block columns of $\mathbf{B}(\scr{C}) $ and
 $\mathbf{C}(\bar{\scr{C}})$  respectively; as will become clear in a moment, the properties which characterize  completeness do not depend
 on these orderings.

For a given  complementary subsystem $( \mathbf{C}(\bar{\scr{C}}),
A,\mathbf{B}(\scr{C}))$ to be complete, its transfer matrix
 $\mathbf{C}(\bar{\scr{C}})(sI-{A})^{-1} \mathbf{B}(\scr{C}) $   must be nonzero and the {\em matrix pencil}
\eq{\pi(\scr{C}) = \matt{\lambda I - {A} & \mathbf{B}(\scr{C})\cr
\mathbf{C}(\bar{\scr{C}}) &0}\label{pen}} must have rank no less
than  $n$  for all real and complex $\lambda $ \{See \cite{ander} or
Corollary 4 of \cite{corf}\}. The requirement that the transfer
matrix of each complementary subsystem be nonzero, can be
established in terms of the connectivity of the ``graph'' of
\rep{generic}. By the {\em graph} of \rep{generic}, written
$\mathbb{G}$, is meant that $k$-vertex  directed graph
    with labels in $\scr{I}$  and arcs defined so that there is an arc from vertex $j$ to $i$
 if $C_{i}(sI-A)^{-1}B_j\neq 0$ for all labels $i,j\in\scr{I}$.  For the transfer matrices of all
  complementary subsystems of \rep{generic} to be nonzero, it is necessary and sufficient that $\mathbb{G}$ be a strongly connected graph
 \{Lemma 8, \cite{corfmat}\}.

\section{Analysis}\label{sec:analysis}

The aim of this section is to prove Proposition \ref{prop1}. To do
this it is useful to first
 establish certain properties of  the sub-system of \rep{decent},
 \rep{decent2} defined by the equations
\begin{eqnarray}\dot{\epsilon} & = &\tilde{A}\epsilon + \sum_{i\in\mathbf{m}}\sum_{j\in\bar{\scr{N}}_i}B_iu_{ij}\label{ndecent}\\
y_{ij}& =&
C_{ij}\epsilon,\;\;\;\;ij\in\scr{J}\label{ndecent2}\end{eqnarray}
where  $\scr{J}$ is the complement of the set
$\{ii:i\in\mathbf{m}\}$ in $\scr{I}$ and for $i\in\mathbf{m}$,
 $\bar{\scr{N}}_i$ is  the complement of  the set $\{i\}$ in $\scr{N}_i$. This sub-system is what results when outputs
  $y_{ii},\;i\in\mathbf{m}$, are deleted from
  \rep{decent2}.
Our  goal here is to show that with suitable scalars $f_{ij}$, the
matrix pairs $(\bar{H},B_p),\;p\in\mathbf{m}$,  are all controllable
with controllability index $m$ where \eq{\bar{H} = \tilde{A}+
\sum_{i\in\mathbf{m}}\sum_{j\in\bar{\scr{N}}_i}B_iF_{ij}C_{ij}\label{nH}}
and $F_{ij} = f_{ij}I_n$. Note that for any $f_{ij}$ and any
$p\in\mathbf{m}$ the submatrix $\matt{B_p &\bar{H}B_p & \cdots  &
\bar{H}^{m-1} B_p}$ has exactly $nm$ columns.
 Since $nm$ is the dimension of the system \rep{ndecent}, \rep{ndecent2},
  $m$ is the smallest possible controllability
index which the pair  $(\bar{H},B_p)$ might attain as the $f_{ij}$
range over all possible values. From this it is obvious that if for
each $p\in\mathbf{m}$, there exist $f_{ij}$ for which
$(\bar{H},B_p)$ has controllability index $m$, then there must be
$f_{ij}$ for which $(\bar{H},B_p)$ has controllability index $m$ for
all $p\in\mathbf{m}$, and moreover the set of $f_{ij}$ for which
this is true is the complement of a proper algebraic set in the
linear space in which the vector of $f_{ij}$ takes values.

To proceed we will first show that with the $f_{ij}$ chosen
properly, the matrix pair $(F,b_m)$ is controllable, where  $F$ is
the $m\times m$ matrix \eq{F =
\sum_{i\in\mathbf{m}}\sum_{j\in\bar{\scr{N}}_i}b_if_{ij}c_{ij}\label{nnnH}}
and for $i\in\mathbf{m}$,  $b_i$ is the $i$th unit vector in $\R^m$.
Note that $F$ is what results when the feedback laws $v_{ij} =
f_{ij}w_{ij} $ are applied to the system
\begin{eqnarray}\dot{z} & = & \sum_{i\in\mathbf{m}}\sum_{j\in\bar{\scr{N}}_i}b_iv_{ij}\label{nd}\\
w_{ij}& =& c_{ij}z,\;\;\;\;ij\in\scr{J}\label{nd2}\end{eqnarray}
where as before, $c_{ij}$
 is the row in the transpose of the incidence matrix of $\mathbb{N}$ corresponding to the arc  from $j$ to $i$.
Note that \rep{nd}, \rep{nd2} can be viewed as  a $m^*$  channel
system where $m^*$ is the number of labels in $\scr{J}$. In view of
the fact that  span$\{b_1,b_2,\ldots,b_m\}=\R^m$, it is obvious that
\rep{nd} is jointly controllable. Let
 $\mathbb{G}$ denote that $m^*$-vertex  directed graph
    with vertex labels in $\scr{J}$  and arcs defined  so that there is an arc from vertex $ij$ to $kq$
 if $c_{kq}(sI)^{-1}b_i\neq 0$ for $j\in \bar{\scr{N}}_i$.

\begin{lemma} If the neighbor graph $\mathbb{N}$ is strongly connected, then $\mathbb{G}$ is strongly connected.
\label{sc} \end{lemma}

\noindent{\bf Proof of Lemma \ref{sc}:} Note that   for each
$j\in\bar{\scr{N}}_i$ ,
 $c_{ij}(s)^{-1}b_i = -\frac{1}{s} $ and
 $c_{ij}(s)^{-1}b_j = \frac{1}{s}$. From these expressions it follows that
   $c_{ij}(sI)^{-1}b_i\neq 0$  and   $c_{ij}(sI)^{-1}b_j\neq 0$ for $i\in\mathbf{m},\;j\in\bar{\scr{N}}_i$. Therefore
  for each $i\in\mathbf{m}$, the subgraph $\mathbb{G}_i$ induced by vertices $ij,\;j\in\bar{\scr{N}}_i$ is complete.
 By the {\em quotient graph} of $\mathbb{G}$, written $\mathbb{Q}$,  is meant that directed graph with  $m$ vertices labeled $1,2,\ldots,m $
 and an arc from $i$ to $k$
 if there is an arc in $\mathbb{G}$ from a vertex in the set  $\{ij:j\in\bar{\scr{N}}_i\}$ to a vertex in the set $\{kq:q\in\bar{\scr{N}}_k\}$.
  Because each of the subgraphs $\mathbb{G}_i$ is complete, $\mathbb{G}$ will be strongly connected if $\mathbb{Q}$ is strongly
  connected. But   $\mathbb{Q} = \mathbb{N}$ so $\mathbb{Q}$ is strongly connected.
 Therefore    $\mathbb{G}$ is strongly connected. $\qed $

\begin{lemma} If the neighbor graph $\mathbb{N}$ is strongly connected, then each complementary subsystem  of \rep{nd} \rep{nd2}
is complete. \label{yyy}\end{lemma}

\noindent{\bf Proof of Lemma \ref{yyy}:} Let $\scr{C}\subset\scr{J}$
be a nonempty subset  and let
 $(\mathbf{C},0_{m\times m}, \mathbf{B})$ be the coefficient matrices of the complementary subsystem determined by $\scr{C}$.
Thus
 $\mathbf{B} = {\rm block\;row}\{b_i:ij\in\scr{C}\}$, and
 $\mathbf{C} = {\rm block\;column}\{c_{ij}:ij\in\bar{\scr{C}}\}$
where $\bar{\scr{C}}$ is the complement of $\scr{C}$ in $\scr{J}$.
To prove the lemma, it is enough to show that the coefficient matrix
triple $(\mathbf{C},0_{m\times m}, \mathbf{B})$ is complete. To
establish completeness
  the transfer matrix
 $\mathbf{C}(sI)^{-1} \mathbf{B} $ must be nonzero and the     matrix pencil
\eq{\pi(\scr{C}) = \matt{\lambda I &\mathbf{B}\cr \mathbf{C} &
0}\label{pencil}} must have rank no less than  $m$ for all real and
complex $\lambda $ \{cf, Corollary 4, \cite{corf}\}.
 In view of Lemma \ref{sc}
  and the assumption that $\mathbb{N}$ is strongly connected, $\mathbb{G}$ is strongly connected. Therefore by Lemma 8 of \cite{corfmat},
  $\mathbf{C}(sI)^{-1}\mathbf{B} \neq 0$.

   To complete the proof it is enough to show that for all
  complex numbers $\lambda $, $\rank \pi(\scr{C}) \geq m $.
In view of the structure of $\pi(\scr{C})$ in \rep{pencil},  it is
clear that for all such $\lambda$, $\rank \pi(\scr{C})\geq \rank
\mathbf{C}   + \rank \mathbf{B}$. To establish completeness, it is
therefore sufficient to show that \eq{\rank \mathbf{C}   + \rank
\mathbf{B}\geq m\label{suff}}

  Let $q\in\mathbf{m}$ denote the number of distinct integers $i$ such that $ij\in\scr{C}$. In view of the definition
   of $\mathbf{B}$,   $\rank \mathbf{B} = q$.  If $q=m$,
  $\rank \mathbf{B} = m$  and \rep{suff} holds.
 Suppose next that $q<m$.  Let $\mathbf{C}^*$ denote the submatrix of $\mathbf{C}$ which results when all
  rows $c_{ij} $ in $\mathbf{C}$ for which
    $ik\in\scr{C}$ for some $k$, are deleted.
  Since $\rank \mathbf{C}\geq \rank \mathbf{C}^*$ and $\rank B = q$, \rep{suff} will hold if
  \eq{\rank \mathbf{C}^* \geq (m-q)\label{doop}}
Corresponding to the definition of $\mathbf{C}^*$, let
$\mathbb{N}^*$ denote the spanning subgraph of $\mathbb{N}$
  which results when  any arc in $\mathbb{N}$ from $i$ to $j$ for which there
   is a $k$ such that $ik\in\scr{C}$ is removed. There are exactly $q$ distinct values of $i$ for which $ik\in\scr{C}$ for some $k$.
   Moreover,  for any such $i$ the corresponding vertex in
$\mathbb{N}^*$ cannot have any outgoing arcs. Since $\mathbb{N}$ is
strongly connected, any other vertex $k$
 in $\mathbb{N}^*$ must have
 at least one outgoing arc not incident on vertex $k$. This means that the un-oriented version
 of $\mathbb{N}^*$ must have at most $q$
 connected components. Thus if $ M_{\mathbb{N}^*}$ is the incidence matrix of
 $\mathbb{N}^*$, then as a consequence of Theorem 8.3.1 of \cite{graph},
\eq{\rank M_{\mathbb{N}^*} \geq m-q\label{pluto}} But  for any
$ij\in\scr{J}$ such that  $ik\not\in\scr{C}$ for some $k$,  $c_{ij}$
is  the row in the transpose of the incidence matrix of
$\mathbb{N}^*$ corresponding to the arc  from $j$ to $i$. Therefore,
up to a possible re-ordering of   rows, $\mathbf{C}^*
=M'_{\mathbb{N}^*}$. From this and \rep{pluto} it follows that
\rep{doop} holds.
 Therefore the lemma is true.
$\qed$

\begin{lemma} Let $A_{n\times n}$, $F_{m\times m }$ and $g_{m\times 1}$ be any given real-values matrices.
There is a $mn\times mn$  nonsingular matrix $T$ such that {\small
\eq{ [G\; HG \; \cdots \;H^{m-1}G]=
      [g\otimes I_n \; (Fg)\otimes I_n \;\cdots \;(F^{m-1}g)\otimes I_n] T \label{lilis1}
}} where $G = g\otimes I_n$ and $H = I_m\otimes A + F\otimes I_n$.
\label{wang}\end{lemma}

\noindent{\bf Proof of Lemma \ref{wang}:} Since $(I_m\otimes
A)(F\otimes I_n)=(F\otimes I_n)(I_n\otimes A)$, for $k\geq 1$
 \begin{eqnarray*}H^k & = & (I_m\otimes A+F\otimes I_n)^k
\\
 & = & \sum_{i=0}^k {k\choose i} F^i\otimes A^{k-i} \end{eqnarray*}
where ${k\choose i}$ is the binomial coefficient. Thus
 \begin{eqnarray} \nonumber H^kG & = & (I_m\otimes A+F\otimes I_n)^k (g\otimes I_n) \\ & = & \sum_{i=0}^k{k\choose i}F^ig\otimes A^{k-i},\;\;\;k\geq 1\label{pip}
 \end{eqnarray}

Define $T_1 = I_{mn}$ and for $k\in\{2,3,\ldots,m\}$ let $T_k$ be
that $mn\times mn$  matrix  composed of $m^2$ $n \times n$
submatrices $T_{ij}(k)$ defined so that $T_{ii}(k) = I_n,
i\in\mathbf{m}$, $T_{(i+1),(k)}(k)={k-1\choose
i}A^{k-i-1},\;i\in\{0,1,\ldots, k-1\}$, and all remaining $T_{ij}(k)
= 0$.

Let $X(k)=[g\otimes I_n \; \cdots \; (F^{k-1}g)\otimes I_n\; H^kG \;
\cdots H^{m-1}G]$ for $k\in \mathbf{m}$. Obviously, $X(1)=\matt{G &
HG &\cdots &H^{m-1}G}$, and $X(m)=\matt{g\otimes I_n & (Fg)\otimes
I_n &\cdots & (F^{m-1}g)\otimes I_n}$.

The definition of $T_k$ and \rep{pip} imply that \eq{ X(k) T_k =
 X(k-1),\; k\geq 1.
\label{xk}} We claim that $T \dfb T_mT_{m-1}\cdots T_{1}$ has the
required properties.  Note first that each of the $T_i$ is an upper
triangular matrix with ones on the main diagonal. Thus each $T_i$ is
nonsingular which implies that $T$ is nonsingular.
 According to \rep{xk}
 \begin{eqnarray*}   [
 g\otimes I_n \; (Fg)\otimes I_n \; & \cdots\; & (F^{m-1}g)\otimes I_n
 ] T \\ & = &  X(m) T_m
  T_{m-1}\cdots T_{1}
  \\ & =&
X(m-1)T_{m-1}  T_{m-2}\cdots T_{1}
\\ & &
\vdots
\\& = &
X(1)T_1.
\end{eqnarray*}
Since $T_1 = I_{mn}$,  \rep{lilis1} is true. $\qed $

\begin{lemma} Suppose $\mathbb{N}$ is strongly connected. The $m^*+m$ channel system  \rep{decent}, \rep{decent2}
is jointly controllable and jointly observable. \label{joint}
\end{lemma}

\noindent{\bf Proof of Lemma \ref{joint}:} In view of the
definitions of the $B_i$,  it is clear that
$\scr{B}_1+\scr{B}_2+\cdots \scr{B}_m = \R^{nm}$ where $\scr{B}_i$
is the column span of $B_i$. It follows at once that \rep{decent},
\rep{decent2} is jointly controllable. To establish joint
observability it is enough to show that $0$ is the only vector
$x\in\R^{nm}$ for which
 $C_{ij}x = 0,\;ij\in\scr{I}$ and   $\tilde{A}x = \lambda x$ for some complex number $\lambda $. Suppose $\tilde{A}x = \lambda x$ in which case
 $Ax_i = \lambda x_i$ where $x = \matt{x_1' &x_2' &\cdots &x_m'}'$ and $x_i\in\R^n,\;i\in\mathbf{m}$. Moreover, if  $C_{ij}x = 0,\;ij\in\scr{I}$,
  then $C_ix_i = 0,\;i\in\mathbf{m}$ and $M_Ix=0$ where $M_I$ is the transpose of the incidence matrix of $\mathbb{N}$.
  Since $\mathbb{N}$ is strongly connected, $M_Ix=0$ implies that $x_i = x_1,\;i\in\mathbf{m}$. Thus $C_ix_1 = 0,\;i\in\mathbf{m}$. But
  $(C,A)$ is observable by assumption where $C=\matt{C_1' &C_2'&\cdots &C_m'}'$.  Therefore $x_1=0$.  This implies that $x=0$ and thus that
\rep{decent}, \rep{decent2} is jointly observable. $\qed$

\noindent{\bf Proof of Proposition \ref{prop1}:} Since span
$\{b_1,b_2,\ldots, b_m\} = \R^m$,  the subsystem defined by \rep{nd}
\rep{nd2} is jointly controllable. From this, Lemma \ref{yyy} and
Theorem 1 of \cite{corfmat} it follows that for each
$p\in\mathbf{m}$, there exist $f_{ij}$ such that $(F, b_p) $ is a
controllable pair where $F$ is as defined \rep{nnnH}. Since the set
of $f_{ij}$ for which this is true, is the complement of  a proper
algebraic set in the space in which the $f_{ij}$ takes values, there
also exist $f_{ij}$ for which $(F, b_p) $ is a controllable pair for
all $p\in\mathbf{m}$. Fix such a set of $f_{ij}$.

By definition $B_i = b_i\otimes I_n \;i\in\mathbf{m}$, $C_{ij} =
c_{ij}\otimes I_n,\;ij\in\scr{J}$ and $\tilde{A} = I_m\otimes A$. In
view of the definition of $\bar{H}$ in \rep{nH}, $\bar{H} =
I_m\otimes A +F\otimes I_n$.
 From this and Lemma \ref{wang} it follows that for each $p\in\mathbf{m}$ there is a nonsingular matrix $T_p$ such that
  $\matt{B_p& \bar{H}B_p & \cdots&  \bar{H}^{m-1}B_p} = (\matt{b_p&  Fb_p & \cdots&  F^{m-1}b_p}\otimes I_n)
  T_p$.
Since  each $T_p$ is nonsingular and each $(F,b_p)$ is a
controllable pair, \[\rank\matt{B_p & \bar{H}B_p &\cdots
&\bar{H}^{m-1}B_p} = nm\] Therefore for each $p\in\mathbf{m}$,
$(\bar{H},B_p)$ is a controllable pair with controllability index
$m$. Note that if we define $F_{ii} = 0,\;i\in\mathbf{m}$, then  in
view of \rep{H}, $H = \bar{H}$. Therefore, for each
$p\in\mathbf{m}$, $(H,B_p)$ is a controllable pair with
controllability index $m$. Clearly this must be true generically,
for almost all $F_{ij},ij\in\scr{I}$.

 In view of Theorem 1 of \cite{corfmat}, the complementary subsystems
 of \rep{decent} and \rep{decent2} must all be complete. But by Lemma \ref{joint}, \rep{decent} and \rep{decent2} is a
 jointly controllable, jointly observable system. From this and Corollary 1 of \cite{corfmat},
it follows that there exist $F_{ij},\; ij\in \scr{I}$ such that for
  all $p\in\mathbf{m}$ and all  $q\in\scr{N}_p$, the matrix pairs $(H,B_p)$  and $(C_{pq},H)$ controllable and observable respectively.
Since this also must be true generically  for almost all $F_{ij}$
the proposition is true. $\qed$

\section{Concluding Remarks}


In this paper we have explained how to construct a family of
distributed observers for a given neighbor graph $\mathbb{N}$ which
are capable of estimating the state of the system \rep{sys1}
\rep{sys2} at an pre-assigned but arbitrarily fast convergence rate.
 There are many  additional issues to be addressed. For example, how might one
 construct distributed observers of least dimension which can estimate $x$? Accomplishing this will almost
 certainly require the transmission to each agent $i$ from each  neighbors $j$,  the signal  $y_j$ which agent $j$ measures. This of course
 comes at a price, so there is a trade-off to be studied between required observer dimension on the  one hand and the amount of information
 to be transferred across the network on the other. Another issue of importance would be to try to construct  a distributed observer
 for the case when $\mathbb{N}$ changes over time; of course this problem will call for a  different type of mathematics since the equations
 involved will be time-varying systems. Finally it would be useful to
 try to determine how to construct
  distributed observers when in place of \rep{sys2},
  one has $\dot{x} = Ax +\sum_{i=1}^mB_iu_i$ where $u_i$ is an input signal which can be measured by agent $i$.
  Some of these problems will be addressed in the future.

\section{Acknowledgement} The authors wish to thank Shinkyu Park and   Nuno C. Martins for
 useful discussions which have contributed to this work.

\bibliographystyle{unsrt}
\bibliography{my,steve}

\begin{IEEEbiography}[{\includegraphics[width=1in,height=1.25in,clip,keepaspectratio]{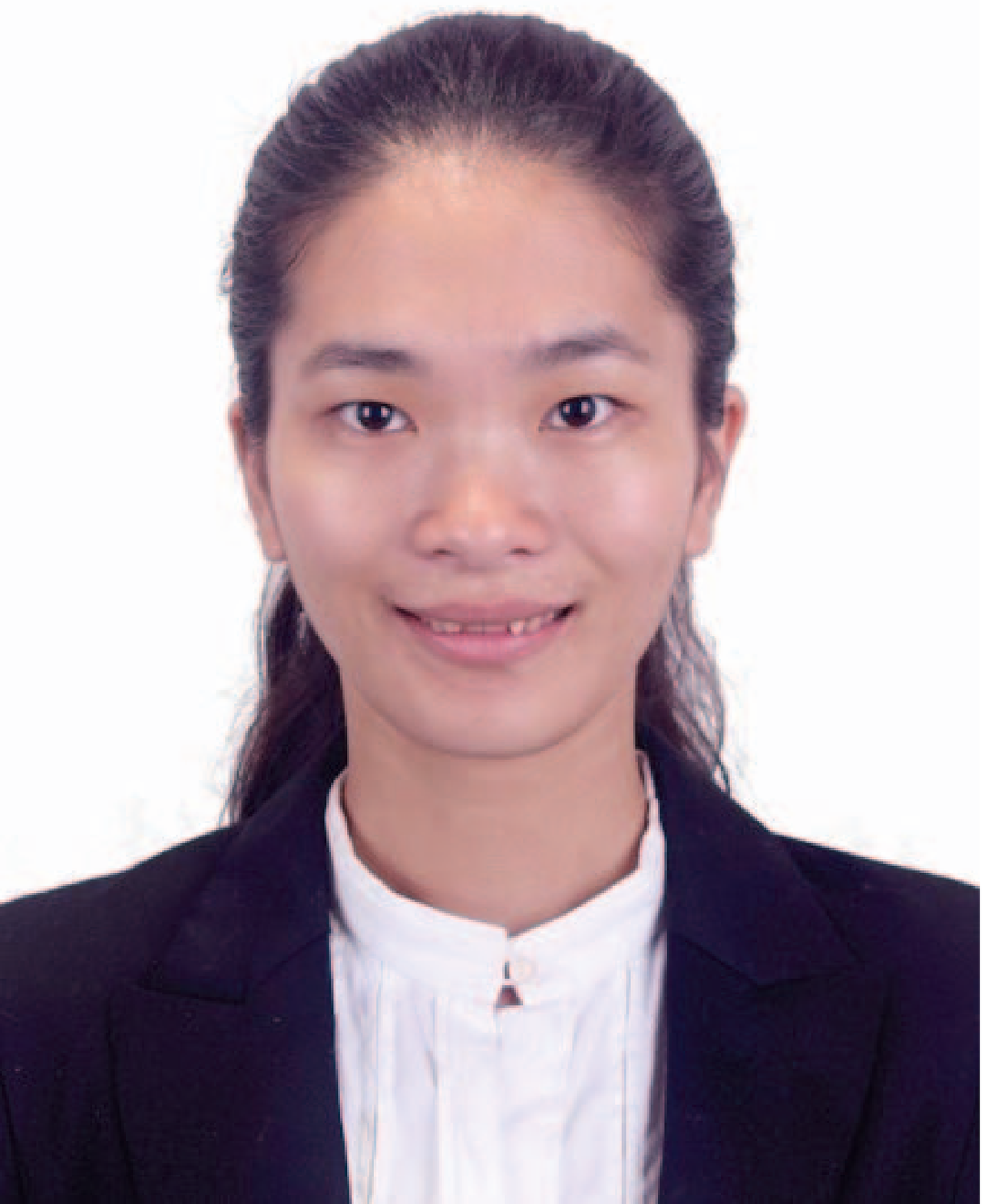}}]{Lili Wang} received the
 B.E. and M.S. degrees from Zhejiang University, Zhejiang, China, in 2011 and 2014, respectively. She is currently a Ph.D. student
 majored in  electrical engineering in  the School of Engineering \& Applied Science, Yale University, USA.  Her research is on
 the topic of cooperative multi-agent systems and distributed observer.
\end{IEEEbiography}

\begin{IEEEbiography}[{\includegraphics[width=1in,height=1.25in,clip,keepaspectratio]{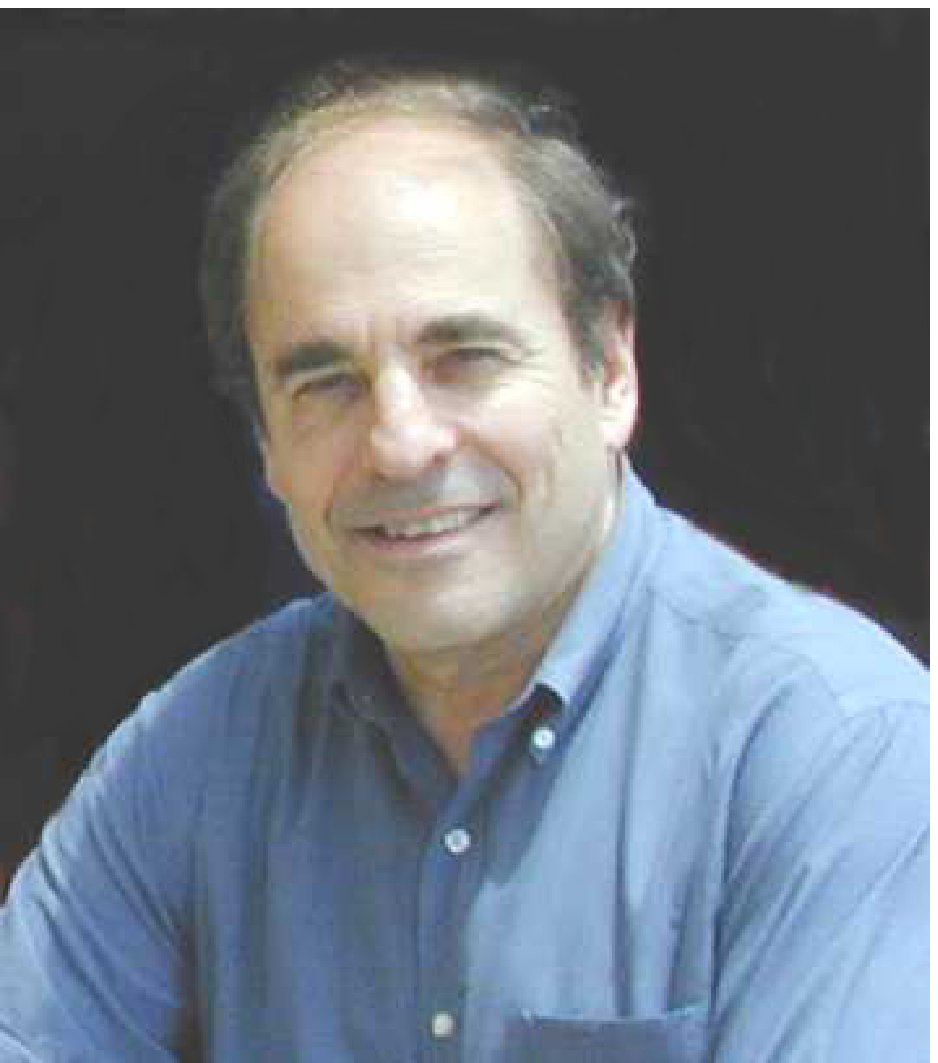}}]{A Stephen Morse}  received a
Ph.D. degree from Purdue University. Since 1970 he has been with
Yale University where he is presently the Dudley Professor of
Engineering. He has received several awards including the 1999 IEEE
Technical Field
 Award for Control Systems and the American Automatic Control Council's 2013 Richard E. Bellman Control Heritage Award.
  He is a member of the National Academy of Engineering.\end{IEEEbiography}

\end{document}